\documentclass[aps,amsmath,amssymb,footinbib,12pt]{revtex4}
\usepackage{graphicx}
\usepackage{amsmath}
\usepackage{enumitem}
\usepackage{natbib}
\usepackage{float}
\usepackage{verbatim}
\usepackage{mathrsfs} 
\usepackage{color}
\usepackage{bm}

\begin{document}
\baselineskip=25 pt
\title{Keldysh formalism for multiple parallel worlds}
\author{Mohammad Ansari and Yuli V. Nazarov}

\affiliation{Kavli Institute of NanoScience, Delft University of Technology, Lorentzweg 1, NL-2628 CJ, Delft, The Netherlands.}

\begin{abstract}
We present here a compact and self-contained review of recently developed Keldysh formalism for multiple parallel worlds. The formalism has been applied to consistent quantum evaluation of the flows of informational quantities, in particular, to evaluation of Renyi and Shannon entropy flows. We start with the formulation of standard and extended Keldysh technique in  single world in a form convenient for our presentation. We explain the use of Keldysh contours encompassing multiple parallel worlds    
In the end, we shortly summarize the concrete results obtained with the method.
\end{abstract}

\maketitle

\section{Introduction}
The seminal work of Leonid Keldysh \cite{Keldysh} has paved the way to modern understanding of quantum systems out of equilibrium. One can do much work in the area armed just with Fermi Golden Rule and some defiance, yet a consistently scientific approach will almost necessary involve the Keldysh formalism. The formalism has been successfully applied for derivations of dynamical equations of complex  systems where intuition ceases to work, like superconductors \cite{LarkinOvchinnikov}, strongly correlated systems \cite{Mahan}, non-linear sigma-models \cite{Kamenev}.     

For many years the formalism was considered too much complicated for a practical researcher and hardly applied beyond several specific fields. ''Keldysh approach'' sounded as a synonym of unnecessary theorization and an antonym to clear physical reasoning.
One of us (Y.N) remembers a talk given by a high-class theorist with a taste for abstract models, young experimentalists being his primary audience. Somewhere in the middle of the talk he said: ''Now let us move to physical quantities, namely, Keldysh Green functions''. A burst of laugh lasted for more than five minutes.

The situation begun to change in nineties, and is quite different nowadays. The formalism receives much more practical attention, more and more theorists and numerical researchers become qualified, nice modern reviews \cite{Jauho,KamenevReview} have appeared in addition to the classical ones \cite{Rammer}.
A unique property of Keldysh formalism that distinguishes it from all other diagrammatic techniques \cite{AGD} is that the zero-order approximation is generally unstable with respect to perturbative corrections. This property is widely appreciated now and makes the technique an indispensable tool for complex quantum dynamics.

Recent extensions of Keldysh technique to non-unitary evolution of density matrix \cite{QT,Kindermann} allow to access non-trivial problems of quantum statistics and analyze large deviations from equilibrium \cite{Tero}. Finite-element approach to Keldysh Green functions for electrons, so-called quantum circuit theory \cite{Circui,QT} proved to be useful to built adequate models of quantum nanostructures. 

Keldysh technique permits natural formulation in terms of path integrals \cite{KamenevReview} providing a very instructive picture of ''doubling'' a classical stochastic variable when put on the Keldysh contour. 
This provides a fundamental link between Keldysh and Feymann-Vernon formalism. The Keldysh action arising in this context can be evaluated by blocks, each block being obtained from a non-unitary evolution \cite{Snyman}. The Landauer-Buttiker \cite{LB1,LB2} scattering approach receives a compact and general formulation in terms of Keldysh action \cite{Snyman,NazarovForButtiker}

All these extensions are still based on time evolution along a single ``doubled'' Keldysh contour. In this paper, we discuss a recent extension in a different direction. Technically, the extension involves time evolution along many ``doubled'' contours. We will refer to these pairs of contours as parallel worlds (this terminology has nothing to do with an attempt of interpretation of quantum mechanics involving parallel worlds). The closing of the contours is typically different for different sub-parts of the quantum system under consideration: for some, the contours are closed separately within each world, while for others they can go back and forth through all the worlds.

As we show below, this formalism is natural and indispensable for evaluating the quantities that are non-linear in the density matrix. The physical meaning of such quantities is not obvious since they do not conform the standard definition of a physical observable, although they are commonly used in quantum information theory \cite{QuantumInformation}, for instance, for entanglement characterization. Most work and applications have been done for evaluation of Renyi entropies \cite{Nazarov11, AN14, AN15}. We follow these papers in our outlining.
    
The paper is organized as follows. 
In Section \ref{standard} we formulate the standard Keldysh formalism in a way convenient for further presentation, putting emphasis on the link between Keldysh technique and master or Bloch equations.
In Section \ref{extended} we explain the extension of the formalism on non-unitary evolution mostly concentrating on the example of full counting statistics of energy flows \cite{Pilgrim} that is useful in the context of Renyi entropy flows.
Then we explain the use of parallel worlds concept for evaluation of the conserving quantities related to the products of density matrices of sub-parts (Section \ref{why}) and formulate the diagrammatic technique for this situation in Section \ref{dia}. The relations between different Keldysh correlators for a (sub)system in thermal equilibrium, so-called KMS \cite{KMS} relations, are important for single-world techniques. We discuss its generalization on multiple worlds in Section \ref{sec:KMS}.  

The rest of the paper is devoted to specific examples for which the general theory can be simplified and elaborated. We concentrate on second-order diagrams in Section \ref{secondorder} and explain the specifics of higher-order diagrams in Section \ref{higherorder}. We shortly review our recent results on quantum heat engines in Section \ref{QHE}. In Section \ref{exact} we discuss a rather general correspondence between the statistics of the energy flows and Renyi entropy flows.
We conclude in Section \ref{conclusions}.

\section{Standard Keldysh formalism}
\label{standard}
Let us first formulate the standard Keldysh formalism in a way that illustrates its potential and at the same time makes direct connections with the problems to be considered further in the text. The starting point of the formalism 
is the formal expression for unitary time evolution of the density matrix $\hat{R}$ of a quantum system governed by a (generally time-dependent) Hamiltonian $\hat{H}(t)$,
\begin{equation}
\hat{R}(t)  = {\rm Texp}\left(i\int^t_{t'} d\tau \hat{H}(\tau)\right) \hat{R}(t')\  {\rm \tilde{T}exp}\left(-i\int_{t'}^t d\tau \hat{H}(\tau)\right)
\label{eq:QuantumEvolution}
\end{equation}
${\rm Texp}({\rm \tilde{T}exp})$ denote time(anti)ordering in the evolution exponents. 

If we were up to exact quantum evolution of the whole system, we would not need any Keldysh technique: a Schr\''{o}dinder equation would suffice. At the same time, the resulting density matrix would keep the memory of the initial one for infinite time. This is rather unphysical. To address physical situations, we need to separate quantum variables onto relevant and less relevant ones. Quite generally, it can be achieved by {\it bipartition} of the Hilbert space: we present it as a direct product $A\otimes B$ of two subparts $A$ and $B$. The Hamiltonian is separated as
\begin{equation}
\hat{H} = \hat{H}_A +\hat{H}_B +\hat{H}_{AB},
\end{equation}
$\hat{H}_{AB}$ being an operator that involves degrees of freedom in both subspaces, while $H_{A,B}$ work in their respective partitions only.

This opens up the opportunities to treat a great variety of physical situations. For instance, the system $A$ can be a small system with finite number of states while $B$ can be an environment with infinite number of degrees of freedom. In this case, the density matrix of $B$ can be regarded as unchangeable in the course of evolution and will play a role of (thermal) reservoir for $A$: the density matrix $R_A$ will try to adjust to the reservoirs. Alternatively, $B$ can be a collection of independent reservoirs kept at different conditions(like temperatures and chemical potentials): the system $A$ will try to adjust to these competing reservoirs providing the flows of physical quantities, e.g. charge or heat, between the reservoirs. Yet another possibility is that  $A$ and $B$ are both reservoirs and $H_{AB}$ representing a junction between the two. In this case, both $R_{A}$ and $R_{B}$ are unchangeable, while the junction provides the flows to both reservoirs.
  
\begin{figure}
\includegraphics[width=0.35\textwidth]{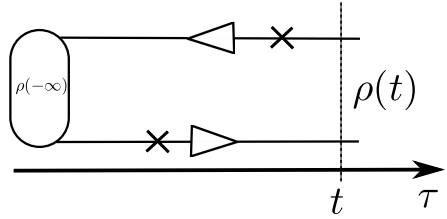}
\caption{ Perturbation theory for a single density matrix on the Keldysh contour.}
\label{fig:commonkeldysh}
\end{figure}

We will assume that the completely separated systems, whose dynamics are governed by $\hat{H}_A +\hat{H}_B$, form a reasonable zero-order approximation and implement a perturbation technique in $\hat{H}_{AB}$ keeping the calculations as general as possible. We assume ``adiabatic switching'' of the perturbation \cite{Landafshiz}:  far in the past the coupling is absent, and the density matrix is a direct product over subspaces $A$ and $B$,
\begin{eqnarray*}
\hat{R}(-\infty) = \hat{R}_A(-\infty) \otimes \hat{R}_B(-\infty) \\
\hat{R}_A(-\infty) = \sum_a p_a |a><a| ;\; \hat{R}_B(-\infty) = \sum_\alpha p_\alpha |\alpha><\alpha|.
\end{eqnarray*}
Here we label the states in subspaces $A$ ($B$) with Latin (Greek) indexes. )
The coupling slowly grows achieving actual values at time long before $t$. The time evolution of the density matrix is given by
\begin{equation}
\label{eq:HABevolution}
\hat{R}(t)  = {\rm Texp}\left(i\int^t_{-\infty} d\tau \hat{H}_{AB}(\tau)\right) \hat{R}(-\infty) {\rm \tilde{T}exp}\left(-i\int_{-\infty}^t d\tau \hat{H}_{AB}(\tau)\right)
\end{equation}
$\hat{H}_{AB}(\tau)$ is taken here in interaction representation.  Expanding this in  $H_{AB}(\tau)$ gives perturbation series most conveniently presented as diagrams involving the Keldysh contour (Fig.\ref{fig:commonkeldysh}). The operators in perturbation series are ordered along the contour. Two parts of the contour correspond to time evolution of bra's and ket's in the density matrix. The crosses represent the (time-dependent) perturbation $H_{AB}(t)$ at a certain time moment. The integration over time moments of all perturbations is implied. There is a state index associated with each piece of the contour. Since $\hat{R}(-\infty)$ is diagonal, this index does not change when passing this element.
The index changes if a non-diagonal matrix element of the perturbation is involved. Summation over the indices is implied.

\begin{figure}
\includegraphics[width=0.45\textwidth]{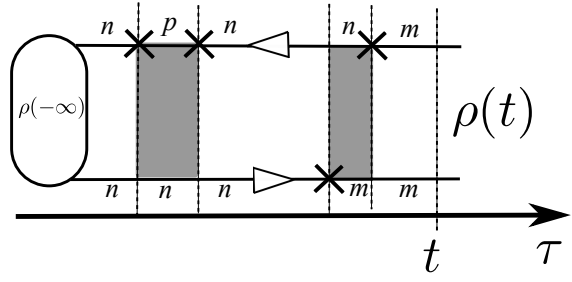}
\caption{ Master equation in Keldysh formalism is obtained from the re-summation of perturbation series whereby the time-line is separated into diagonal and non-diagonal (grey-shaded) blocks. The state index here encompasses the indices in both subspaces. }
\label{fig:masterequation}
\end{figure}

In distinction from most perturbation theories, the zero-order approximation in Keldysh formalism is not stable with respect to small perturbations. For instance, if $A$ is small and $B$ is a reservoir, $\hat{R}_A$ is determined by the reservoir at arbitrary small coupling strength and can have nothing to do with the initial $\hat{R}_A(-\infty)$. This implies that we need to re-sum the perturbation series. 
In a single world, there is a simple way to re-sum the perturbation series and arrive at a master equation that contains only diagonal elements of density matrix (Fig. \ref{fig:masterequation}). For a diagram, we split the time-line by perturbations into the blocks
as shown in the Figure. The blocks come in two sorts: diagonal ones, that have the same state index on both contours,  and non-diagonal
ones. To compute a diagramm, we need to integrate over time duration of each block. For non-diagonal blocks, the integrand is an oscillatory function of time and the integral has a chance to converge. For non-diagonal blocks, the integrand is a constant, and integration diverges. This indicates that the diagrams need to be resumed. If we look at time derivative of the density matrix, it is contributed by the first non-diagonal block. Summation over the subsequent diagonal blocks replaces $\hat{R}(-\infty)$ with the density matrix at time moment right after the first non-diagonal block. With this, the evolution equation for diagonal matrix elements $p_{a \alpha}(t) \equiv R_{a \alpha, a \alpha}$ can be written as (assuming summation over the repeating indices)
\begin{equation}
\frac{d}{dt} p_{a \alpha}(t) = \int_0^{\infty} d\tau W_{a \alpha,b \beta}(\tau) p_{b \beta}(t-\tau).
\label{eq:re-summation1}
\end{equation}
$W_{a \alpha,b \beta}(\tau)$ being the sum of the perturbation expansion comprising a non-diagonal block that starts from the second order in $\hat{H}_{AB}$. 
It is natural to require that the matrix elements  of $H_{AB}$ are only non-diagonal, that is, $H^{(AB)}_{a \alpha, b \beta} = 0$ if either $a=b$ or $\alpha=\beta$.

If $p_{a \alpha}(t)$ changes slowly in comparison with the typical time-scale of the blocks, one can neglect this time-dependence under the integration sign.   The integration over time duration of the blocks with different indices gives the transition rates $\Gamma_{a\alpha,b\beta}=\int_0^{\infty} d\tau W_{a \alpha,b \beta}(\tau)$. The unitarity guarantees that the integration over time duration of the blocks with the same indices gives $\int_0^{\infty} d\tau W_{a \alpha,a \alpha}(\tau)=-\Gamma_{a\alpha,b\beta}$ the total transition rate from the state
$|a\alpha>$, that is the sum of the partial transition rates, $\Gamma_{a\alpha}=\sum_{b\beta} \Gamma_{a\alpha,b\beta}$.    In this way, we come to the master equation in the traditional form,

\begin{equation}
\frac{d}{dt} p_{a \alpha} = - \Gamma_{a\alpha} p_{a \alpha} + \Gamma_{a\alpha,b\beta} p_{b\beta}.
\label{eq:master}
\end{equation}

There are situations when the non-diagonal elements of density matrix are also relevant for dynamics. For instance, a relevant subset of quantum states in $A$ can be approximately degenerate such that their energy separations are of the order of the rates $\Gamma$ or such degeneracy is provided by a coherent drive with a frequency that cancels the energy separations. A generic example is the quantum heat engine described in Section \ref{QHE}. Let us treat the system $B$ as a reservoir for the system $A$ and sum up over its states assuming unchanged $\hat{R}_B$.  Instead of the diagonal blocks, we define the blocks where the states at two parts of the contour belong to the relevant subset for $A$ and the same state for $B$.  After the re-summation over these blocks, the evolution of the density matrix in the relevant subset is given by 
\begin{equation}
\frac{d}{dt} \rho_{ab} = i \left(H^r_{ac}\rho_{cb} - \rho_{ac} H^r_{cb}\right) +\int_0^\infty W_{ab,cd}(\tau) \rho_{cd}(t-\tau).
\label{eq:re-summation2} 
\end{equation}
Here $\hat{H}^r$ is an operator accounting for a weak degeneracy lifting in the relevant set and typically includes the coherent drive.
Under assumptions of slow change of this matrix, $\rho_{cd}(t-\tau)\approx \rho_{cd}(t)$, one can perform integration over the time duration $\tau$ of the blocks to arrive at 
Bloch equation in its traditional local-in-time form,
\begin{equation}
\frac{d}{dt} \rho_{ab} = i \left(H^r_{ac}\rho_{cb} - \rho_{ac} H^r_{cb}\right) +\Gamma_{ab,cd} \rho_{cd}.
\label{eq:Bloch}
\end{equation}
The common feature of the equations (\ref{eq:re-summation1}),(\ref{eq:master}),(\ref{eq:re-summation2}),(\ref{eq:Bloch}) is the existence of the stationary solution. Mathematically, the linear operator acting on a density matrix in the right hand side of the equations has a zero eigenvalue. The existence of the stationary solution is obvious from physical reasons and is a consequence of unitary dynamics. The system approaches the stationary solution no matter from what initial condition it started its evolution, it forgets the initial conditions. As we will see, the extensions of the Keldysh formalism typically do not have a stationary solution.

\section{Extended Keldysh technique}
\label{extended}

The extended Keldysh technique is formally defined through an evolution with the Hamiltonians $\hat{H}^{+,-}$ that are different at forward and backward part of the Keldysh contour \cite{QT,Kindermann}.
\begin{equation}
\hat{R}(t)  = {\rm Texp}\left(i\int^t_{-\infty} d\tau \hat{H}^+(\tau)\right) \hat{R}(-\infty) {\rm \tilde{T}exp}\left(-i\int_{-\infty}^t d\tau \hat{H}^-(\tau)\right)
\label{eq:ExtendedQuantumEvolution}
\end{equation}
While this equation is very similar to Eq. (\ref{eq:QuantumEvolution}), the evolution for different Hamiltonians is not unitary. Consequently, $\hat{R}(t)$ is not a density matrix, in particular, its trace is not $1$. It is natural to call it a pseudo-density matrix. Apparently, it is not physical: what is the physical use of it?

Let us set the Hamiltonians to $\hat{H}^{\pm}(\tau) = \hat{H}_0 + \chi^{\pm}(\tau) \hat{I}$, and compute the trace ${\rm Tr}[\hat{R}(t)]$. This depends on the values 
of $\chi^{\pm}(\tau)$ for all time moments preceding $t$, ${\rm Tr}[\hat{R}(t)]\equiv \exp({\cal S}\{\chi^{\pm}(\tau)\})$. By expanding Eq. (\ref{eq:ExtendedQuantumEvolution}) in $\chi^{\pm}(\tau)$ one sees that the ${\cal S}\{\chi^{\pm}(\tau)\})$ is nothing but the generating function of all possible Keldysh cumulants of the operator $\hat{I}$ taken at different moments of time. Therefore, it completely characterizes the time-dependet quantum fluctuations. 

The functional ${\cal S}\{\chi^{\pm}(\tau)\})$ is called Keldysh action and is routinely applied in the context of path-integral formulation of the formalism \cite{KamenevReview}.  In this case, $H_0$ describes a sub-system subject to a quantum field $\chi^{\pm}$ that typically arises in the course of path-integral representation of this variable, and ${\cal S}\{\chi^{\pm}(\tau)\})$describes the response and back-action of the sub-system on this field. It can be used as a block in Feymann-Vernon action that describes the fluctuation dynamics of the field \cite{Snyman}.

Another application of the extended Keldysh formalism is the full counting statistics (FCS) \cite{Kindermann}. Let us set $\chi^\pm = - \chi/2$, $\chi$ being a constant in time interval $(0,{\cal T})$ and is called counting field. The expansion of the Keldysh action in $\chi$ produces the Keldysh-time-ordered cumulants of the quantum variable $Q = \int_0^{\cal T} d\tau I(\tau)$. Under certain conditions \cite{Kindermann}, the inverse of this generating function gives the probability of a change $Q$ of this variable during the time interval,
\begin{equation}
P(Q) = \int d\chi e^{i \chi Q} e^{{\cal S}\chi)}.
\end{equation}   
This technique has been implemented for FCS of the charge transferred between the reservoirs. \cite{Levitov,QT,Kindermann}

An accurate definition of FCS for conserving quantities implements a gauge transform in a bipartition. Let us consider an operator of a conserving quantity $\hat{O}$ that is separable in the bipartition, $\hat{O} = \hat{O}_A + \hat{O}_B$. Let us define a unitary transformation $\hat{U_{A}}(\chi) = \exp(i \chi\hat{O}_A)$ and the Hamiltonians on two parts of the contour as
\begin{equation}
\hat{H}^{\pm} = \hat{U_{A}}(\pm\chi/2)\  \hat{H} \  \hat{U_{A}}(\mp\chi/2).
\end{equation}
Since $H_{A,B}$ commute with $\hat{O}$, the coupling $H_{AB}$ is the only part modified by this transform,
\begin{equation}
\hat{H}^{\pm} =  \hat{H}_A + \hat{H}_B + \hat{H}^{\pm}_{AB}
\end{equation}
The evolution of the pseudo-density matrix is given by an extension of Eq. (\ref{eq:HABevolution}),
\begin{equation}
\hat{R}(t)  = {\rm Texp}\left(i\int^t_{-\infty} d\tau \hat{H}^{+}_{AB}(\tau)\right) \hat{R}(-\infty)  {\rm \tilde{T}exp}\left(-i\int_{-\infty}^t d\tau \hat{H}^{-}_{AB}(\tau)\right).
\label{eq:HABEvolutionExtended}
\end{equation} 
Trace of $\hat{R}(t)$ defines a Keldysh action ${\cal S}\chi)$ that gives the statistics of transfers of the quantity $\hat{O}$ to/from the subsystem $A$.

This can describe the statistics of conserved quantities such as current and energy flows, the latter is of a special interest for us. In this case, the conserving quantity is the energy $H_A+H_B$ \cite{Pilgrim,Tero}. The unitary transform is equivalent to time-shift of the operators in the interaction representation. The coupling $\hat{H}_{AB}$ can be quite generally presented as a sum of the products of the operators $\hat{A}_i,\hat{B}_i$ working in the corresponding subspaces, $\hat{H}_{AB} = \hat{A}_i\hat{B}_i$. The modified $\hat{H}_{AB}$ then reads 
\begin{equation}
\hat{H}^{\pm}_{AB}(t) = \hat{A}_i(t\mp\chi/2) \hat{B}_i(t).
\end{equation}

The re-summation of the perturbation series made in the previous section is also relevant and shall be made for extended Keldysh formalism. The analogue equations can be derived. Importantly, since the dynamics is non-unitary, the blocks and rates do not satisfy sum rules imposed by unitarity and are generally dependent on counting fields. For instance, in the extended master equation (c.f. with Eq. (\ref{eq:master})),
\begin{equation}
\label{eq:masterExtended}
\frac{d}{dt} p_{a \alpha} = - \tilde{\Gamma}_{a\alpha} p_{a \alpha} + \Gamma_{a\alpha,b\beta} p_{b\beta},
\end{equation}
$\tilde{\Gamma}_{a\alpha} \ne \sum_{b,\beta}\Gamma_{b\beta,
a\alpha}$. 

Owing to this, there is no stationary solution to these equations even for stationary counting fields. The diagonalization of the linear evolution operator gives a set of the solutions of the form 
\begin{equation}
\hat{R}(t) \propto \exp(-D_i t)
\label{eq:eigenvaluesD}
\end{equation}
, $D_i$ being the eigenvalues of the operator. In the long time limit, the general solution will be given by the eigenvalue with the smallest real part,$D_0$.
This gives a remarkably simple and constructive expression for Keldysh action in the limit of long time intervals ${\cal T}$,
\begin{equation}
{\cal S}\chi) = - {\cal T} D_0.
\end{equation}

\section{Why multiple worlds?}
\label{why}
Although this fact is rarely discussed, in addition to physical conserving quantities that are presented by operators there are conserving quantities that are characteristics of a density matrix. They are formally unphysical since they are not associated with any physical operator observable.
An example is provided by the R\'{e}nyi entropies that are defined as traces of integer powers of the density matrix of a closed system$\hat{R}$,
\begin{equation}
S_M = {\rm Tr} \left\lbrace \hat{R}^M\right\rbrace
\label{eq:definition}
\end{equation} 
Since the quantum evolution of the system is governed by a Hamiltonian $\hat{H}$ and
$$
-i \hbar \frac{d \hat{R}}{dt} = [\hat{H},\hat{R}],
$$ 
the density matrices in different momets of time are related by unitary transform and the trace of any power of $\hat{R}$ does not depend on time, $d S_M/dt =0$. 
The definition can be obviously extended to non-integer $M$. The more common Shannon entropy is obtained by taking the limit
\begin{equation}
S = - {\rm Tr} \lbrace \hat{R} \ln \hat R\rbrace = - \lim_{M\to 1} \frac{ \partial S_{M}}{\partial M} = - \lim_{M \to 1} (\ln S_M/(M-1)).
\end{equation}
We note that the $\ln S_M$ is an extensive quantity proportional to the system volume.

Let us now return to the context of bipartition. For two systems $A$ and $B$.
we can now define two sets of R\'{e}nyi entropies,
\begin{equation}
S^{(A)}_M = {\rm Tr}_A  \left\lbrace\left(\hat{R}^{(A)}\right)^M\right\rbrace; \;
S^{(B)}_M = {\rm Tr}_B  \left\lbrace\left(\hat{R}^{(B)}\right)^M\right\rbrace;
\end{equation}
where the reduced density matrices in two subspaces are defined 
via the partial traces in these subspaces,
\begin{equation}
\hat{R}^{(A)} = {\rm Tr}_B \lbrace \hat{R}\rbrace; \;
\hat{R}^{(B)} = {\rm Tr}_A \lbrace \hat{R}\rbrace.
\end{equation}
If the quantum evolutions of the systems are completely independent, 
$$
\hat{H} = \hat{H}_A + \hat{H}_B,
$$
$H_{A,B}$ being operators involving the corresponding subspaces only, the entropies of both sets provide the conserved measures,
$$
\frac{d}{dt} S^{(A)}_M = \frac{d}{dt} S^{(B)}_M=0.
$$

If we take into account the coupling $\hat{H}_AB$, the R\'{e}nyi entropies are not conserved anymore. Let us assume at the moment  the systems $A$ and $B$ are infitely large and are characterized by continuous excitation spectrum while $H_{AB}$  couples a relatively small number of degrees of freedom in both systems. This situation is similar to that of two metallic leads kept at different chemical potentials and containing practically infinite number of electrons. If the leads are connected by a small junction, finite electric current flows through the junction, while the distribution of electrons in infinite leads remains unchanged. From this analogy, it is natural to conjecture that a finite R\'{e}nyi entropy flow, Re-flow, flows between the subsystems $A$ and $B$. We define the flows as  time derivatives of extensive quantities,
\begin{equation}
{\cal F}^{(A),(B)}_M =  \frac{d}{dt} \ln S^{(A),(B)}_M.
\end{equation}  
Owing to conservation of R\'{e}neyi entropy in each system, the Re-flows would not depend on exact bipartition of the system and are determined by properties of the coupling that is in principle described by $\hat{H}_{AB}$ rather then by the properties of the systems $A$ and $B$, in full analogy with electric current.
There is, however, an important difference. For physical quantities the conservation holds in the whole system as well as in each subsystems. For instance, elecrtical currents to each lead must satisfy  $I_A+I_B =0$.
 As far as R\'{e}nyi entropies are concerned, there is no exact conservation law for a sum
 $\ln S^{(A)}_M + \ln S^{(B)}_M $ at finite $\hat{H}_{AB}$, although these quatities are extensive. There is a conservation law for the total R\'{e}nyi entropy $\ln S^{(A+B)}$. However, the latter at finite $\hat{H}_{AB}$ is the sum  $\ln S^{(A)}_M + \ln S^{(B)}_M $ only approximately, up to the terms proportional to the volume of the system. Therefore, in general 
$$ 
{\cal F}^{(A)}_M + {\cal F}^{(B)}_M \ne 0.
$$ 

\begin{figure}
\includegraphics[width=0.35\textwidth]{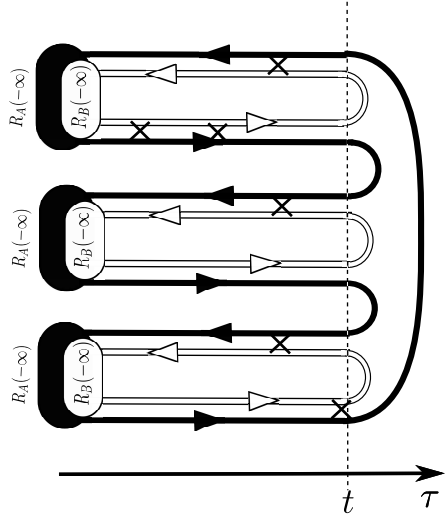}
\caption{ A diagram of perturbation theory for $S^{(A)}_M$ for $M=3$. It involves
three parallel worlds. Reconnection of Keldysh contours for subspaces $A$(black) and $B$(white) accounts for partial trace over $B$ and matrix multiplication in $A$.}
\label{fig:3M}
\end{figure}

How to compute the flows? 
The crucial observation is that the standard Keldysh formalism  as expressed by Eq. (\ref{eq:HABevolution}) can be straightforwardly generalized to any integer number $M$ of density matrices. These matrices undergo independent unitary evolution in time interval $(-\infty,t)$.
It is constructive to think of a set of $M$ ''parallel worlds'' and draw the diagrams for perturbation series  using $M$ parallel bra- and ket-contours.  To compute $S^(A)_M(t)$ with this set, we first need to 
'split' the contours to account for possibly different ordering of operators in subspaces $A$ and $B$ (black and white curves in Fig. \ref{fig:3M} that gives the example for $M=3$). Then we need to {\it reconnect} the contours at  $\tau = t$. All white contours are closed within each world, this corresponds to the partial trace over $B$ for each density matrix involved. In contrast to this, the black contours are connected to form a single loop going through all the worlds, this corresponds to the matrix multiplication in the definition (\ref{eq:definition}) of R\'{e}nyi entropy.  This conveniently represents  the rules of operator ordering for any diagram of particular order in $H_{AB}$.  

\begin{figure}
\includegraphics[width=0.35\textwidth]{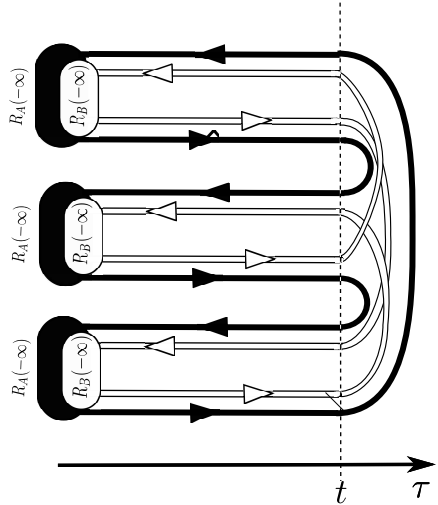}
\caption{  Reconnection of Keldysh contours for conserved measure $K$ defined by Eq. (\ref{eq:other-measure})}
\label{fig:contoursForK}
\end{figure}

It is interesting to note that the sets of R\'{e}nyi entropies are not the only conserved measures characteristic for a bipartition. Any polynomial in density matrix that is invariant with respect to the group $U_A \otimes U_B$ of unitary transforms in two subspaces, would provide such a measure. To give a minimal example, let us label the states in $A$($B$) with Latin (Greek) indices. The quantity 
\begin{equation}
K \equiv  R_{a \alpha,b \gamma} R_{b \beta, c\alpha} R_{c \gamma, a \beta}  
\label{eq:other-measure}
\end{equation}
is a conserved measure that can be reduced neither to the R\'{e}nyi entropies of the systems nor to the R\'{e}nyi entropy of the whole system. It is interesting to note that the reconnecting the contours in a different fashion gives rise to perturbation theory for other conserved measures. For instance, for $K$ the contours are reconnected as shown in Fig. \ref{fig:contoursForK}. The characterization of all such measures forms an interesting research task beyond the scope of this article.

\section{Diagrammatic techniques for multiple worlds}
\label{dia}
Let us illustrate diagrammatic techniques arising in this new context. We concentrate on evaluation of $S_M^{(A)}$ and skip the index $A$ for brevity.
It is natural to require that the matrix elements  of $H_{AB}$ are only non-diagonal, that is, $H^{(AB)}_{a \alpha, b \beta} = 0$ if either $a=b$ or $\alpha=\beta$. 
In this case, the first non vanishing diagram giving the  correction to $S_M$ will be of the second order  in $H_{AB}$. Expressing it in terms of the corrections to $\hat{R}_A$, we find
\begin{equation}
\label{eq:S2M}
\delta S^{(2)}_{M} = M \sum_{0\le N \le M} {\rm Tr}_A\left[\delta \hat{R}^{(1)}_A \hat{R}^N_A \hat{R}^{(1)}_A \hat{R}^{M-N-2}_A \right] + M {\rm Tr}_A\left[ \delta \hat{R}^{(2)}_A \hat{R}^N_A\right].
\end{equation} 
Here we use the symmetry of the parallel worlds cyclically permuting $\hat{R}_A$ and its corrections under trace, this gives the factors $M$ in front of the terms. While there is a  first-order correction to $\hat{R}$, it is non-diagonal in $B$ space. Thus $\delta \hat{R}^{(1)}_A =0$: we will see that it is not the case in the case of quantum heat engine (Section \ref{QHE}) where the non-diagonal elements are important. We only need to deal with $\delta \hat{R}^{(2)}_A$ that is concentrated in a single world. Expansion in $\hat{H}_AB$ gives four terms that correspond to four ways to place two $\hat{H}_{AB}$ on two parts of the contour in a single world,
\begin{eqnarray}
\delta \hat{R}^{(2)}_A = \int_{-\infty}^{t} dt_1 \int_{-\infty}^{t_1} dt_2 \left(  
- \hat{H}_{AB}(t_1) \hat{H}_{AB}(t_2) \hat{R}_A \hat{R}_B
-   \hat{R}_A \hat{R}_B \hat{H}_{AB}(t_2)\hat{H}_{AB}(t_1)
\right.\nonumber \\
\left. + \hat{H}_{AB}(t_1) \hat{R}_A \hat{R}_B\hat{H}_{AB}(t_2) 
+   \hat{H}_{AB}(t_2) \hat{R}_A \hat{R}_B\hat{H}_{AB}(t_1) \right) \label{eq:R2A}
\end{eqnarray} 
We need to substitute this to Eq. (\ref{eq:S2M}). Let us now assume that $\hat{H}_{AB} = \hat{A}_i \hat{B}_i$, $\hat{A}_i$,$\hat{B}_i$ acting on corresponding subspaces. 
Let us introduce the correlators of these operators. Since the contours for the space $B$ are closed within each world, the correlator takes a usual form
\begin{equation}
C_{ij}(t_1,t_2) \equiv {\rm Tr}_B\left[ \hat{B}_i(t_1)\hat{B}_j(t_2) \hat{R}_B\right] 
\label{eq:Ccorrs}
\end{equation} 
A general two-operator correlator in space $A$ is defined as 
\begin{equation}
K^{N,M}_{ij} \equiv {\rm Tr}_A\left[ \hat{A}_i(t_1)\hat{R}^{N}_A\hat{A}_j(t_2) \hat{R}^{M-N}_A\right] S^{-1}_{M}
\label{eq:Kcorrs}
\end{equation}
indices $N,M, 0\le N \le M$ corresponding to different arrangements on the contour traversing $M$ parallel worlds. We divide by $S^{-1}_M$ to keep this correlator
an extensive quantity. 
With this, 
\begin{equation*}
\delta S^{(2)}_M /S_M = \int_{-\infty}^{t} dt_1 \int_{-\infty}^{t_1} dt_2 W(t_1-t_2) = \int_{-\infty}^{t} d t_1\int_0^{\infty} d\tau W(\tau)
\end{equation*}
where the block $W(t_1,t_2)$ is expressed as 
\begin{eqnarray}
W(t_1,t_2) = -C_{ij}(t_1,t_2)K^{0,M}_{ij}(t_1,t_2) - C_{ji}(t_2,t_1)K^{0,M}_{ji}(t_2,t_1) \nonumber \\
 + C_{ji}(t_2,t_1)K^{1,M}_{ij}(t_1,t_2) + C_{ij}(t_1,t_2)K^{0,M}_{ij}(t_1,t_2), \label{eq:Wblock}
\end{eqnarray}
four terms in this equation corresponding to four terms in Eq. (\ref{eq:R2A}). 
So that, the Re-flow is expressed in terms of the block $W$ as 
\begin{equation}
{\cal F}_M = \int_0^{\infty} d\tau W(\tau). 
\end{equation}
More complex diagrams are expressed in terms of those and higher-order correlators that have the similar structure.

Expectedly, the correction to $S_M$ diverges with growing $t$, so that the Keldysh formalism for multiple parallel worlds also requires re-summation. One can introduce one big density matrix $R_{{\bf a} \bm{\alpha}, {\bf b} \bm{ \beta}}$ where the M-dimensional ''vector'' index ${\bf a}$ comprises the state indices in space $A$ for all bra contours, and all other indices are defined similarly. The reduction of this density matrix and re-summation of diagonal blocks leads to the analogues of the Eqs. (\ref{eq:re-summation1}),(\ref{eq:master}),(\ref{eq:re-summation2}),(\ref{eq:Bloch}). For instance, the analogue of Eq. (\ref{eq:re-summation1}) in parallel worlds reads as follows,
\begin{equation}\label{eq:multiplere-summation1}
\frac{d}{dt} p_{{\bf a} \bm{\alpha}}(t) = \int_0^{\infty} d\tau W_{{\bf a} \bm{\alpha},{\bf b} \bm{\beta}}(\tau) p_{{\bf b} \bm{\beta}}(t-\tau).
\end{equation}
$W$ being the blocks computed similarly to Eq. (\ref{eq:Wblock}). Similar to that for extended Keldysh technique (see Eq. (\ref{eq:eigenvaluesD})) , this equation has a set of  non-stationary solutions $\hat{R}(t) \propto \exp(-D_i t)$. The eigenvalues $D_i$ and the form of the solution are affected by the way the contours are re-connected at $t$. For the connection way that gives R\'{e}neyi entropies, the Re-flows are expressed in terms of the eigenvalue with the smallest real part,$D_0$, that depends on the number of the worlds involved,
\begin{equation}
{\cal F}_M = D_0(M).
\end{equation}

\section{KMS relations for multiple worlds}
\label{sec:KMS}
The correlators in a general non-equilibrium system are independent. The state of thermal equilibrium brings about extra relations between the correlators, that are important since they reduce a number of independent parameters in the models of quantum systems. These relations are traditionally called  Kubo-Martin-Schwinger relations \cite{KMS}. For instance,  the correlators $C_{ij}$ (Eq. (\ref{eq:Ccorrs})) in frequency representation are expressed in a KMS state at temperature $T$ in terms of the real part of dynamical susceptibility
$\tilde{\chi}_{ij}(\omega)$ 
\begin{equation}
C_{ij}(\omega) = n_B(\omega) \tilde{\chi}_{ij}(\omega)
\end{equation},
where $n_B(\omega) \equiv 1/(e^{\beta \omega} -1)$, $\beta = \hbar/k_B T$.

Let us show that similar relations hold for the multi-world correlators $K_{ij}(\omega)$ defined by Eq. (\ref{eq:Kcorrs}).

In frequency representation,
\begin{equation*}
K^{N,M}_{ij}\left(\omega\right)  =  \int d \tau e^{i\nu \tau} {\rm Tr} \{ \hat{A}_i(0) \hat{R}_A^N \hat{A}_j(\tau) \hat{R}_A^{M-N} \} / {\rm Tr}{\hat{R}_A^M}
\end{equation*}
 
This correlator  can be rewritten in the energy basis.

\begin{eqnarray}\nonumber
&& K_{i,j}^{N,M}=   \int d \tau e^{i\omega \tau}  \sum_{n,m} \left( A_{i,nm} \frac{e^{-\beta N E_m}}{Z(\beta)^N} A_{j,mn} e^{i (E_m-E_n)\tau} \frac{e^{-\beta E_n (M-N)}}{Z(\beta)^{M-N}} \right) \frac{Z(\beta)^M}{Z(\beta M)} \\  
&&= 2\pi  \delta\left( E_m-E_n+\omega \right) \frac{A_{i,nm}  A_{j,mn} e^{-\beta E_n M} }{Z(\beta M)}e^{\beta N \omega}
\label{eq:corrsinbasis}
\end{eqnarray}
where $Z(\beta)$ is the partition function defined as $Z(\beta)=\sum_i e^{-\beta E_i}$. The standard one-world correlator reads  $K^{0,1}_{ij}\left(\omega\right)  =  \int d \tau \exp({i\omega \tau}) {\rm Tr} \{ \hat{A}_i(0)  \hat{A}_j(\tau) \hat{R}_A \}/ {\rm Tr}{\hat{R}_A}$ which after simplification becomes equal $ 2\pi  \delta\left( E_m-E_n+\nu \right) A_{i,nm}  B_{j,mn} e^{-\beta E_n } /Z(\beta )$. The KMS relation links this to dynamical susceptibility: $K^{0,1}_{ij}(\nu)=\tilde{\chi}_{ij}(\nu)n_B(\nu/T)$. By substituting this in Eq. (\ref{eq:corrsinbasis}) a generalized KMS relation is obtained:
\begin{equation}
K^{N,M}_{ij}\left(\omega\right) =  n_B\left(M\omega\right) e^{\beta\omega N} \tilde{\chi}_{ij}\left(\omega,\beta^*\right)\label{eq:KMS}
\end{equation}
While the correlators are for the system at inverse temperature $\beta$, the dynamical susceptibility is taken at {\it different} inverse temperature $\beta^*\equiv  M \beta$.
Such temperature rescaling looks surprising in the context of KMS relations. However, this is natural in the context of R\'{e}nyi entropies. In the state of thermal equilibrium, the R\'{e}nyi entropy is expressed in terms of free energy at the native and rescaled temperatures,
\begin{equation}
\ln S_M(\beta)= M\beta \left(F(\beta^*) - F(\beta)\right).
\end{equation}

\section{Example: simplicity with second-order diagrams}
\label{secondorder}
Let us start with examples of the multi-world Keldysh approach described. In this Section, we elaborate on second-order diagrams and obtain a rather general picture of Re-flows in this approximation. In a single world, the higher-order diagrams change the values of the rates not changing the dynamics qualitatively. As we will see in the next Section, this is not the case in multiple worlds: there, the higher-order diagrams do bring a qualitative change.

\begin{figure}
\label{fig:second-order}
\includegraphics[width=0.35\textwidth]{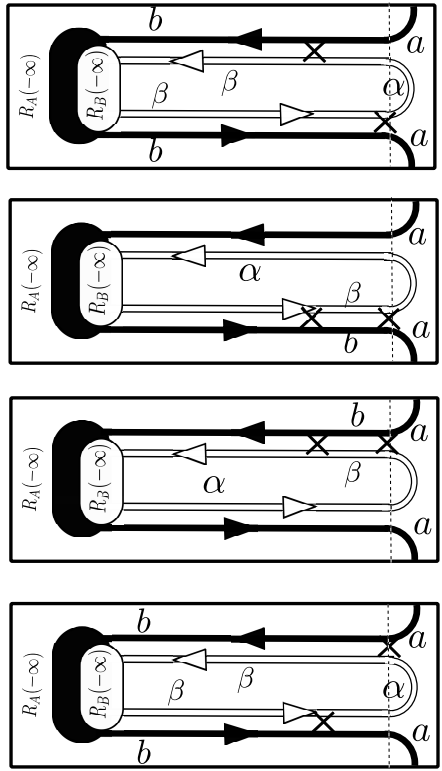}
\caption{ Second order diagrams for time derivative of a R\'{e}nyi entropy.
The contributions come only from perturbations $\hat{H}^{(AB)}$ in
the same world, only this world is shown in each diagram. For all diagrams, the perturbations are taken at time moments $t$ and $t' <t$.
The letters at the contours label the states involved.}
\end{figure}

We compute the Re-flows in the second order in $H_{AB}$ in a way slightly different from that used in the previous Section.
It is proficient to directly compute the time-derivative of $S_M$. For diagrams, this corresponds to placing one of the perturbations at $\tau=t$. The only way to satisfy the continuity of state index along the white contours is to place the second perturbation in the same world. Four contributing diagrams are given in Fig. \ref{fig:second-order}. We notice that the same four diagrams arise in the derivation of Golden Rule transition rate in the standard Keldysh formalism. The specifics of R\'{e}nyi entropies is reflected in extra factors $p^{M-1}_a$ the diagrams acquire in comparison with the case of a single density matrix. 
We do not separate $\hat{H}_{AB}$ into subspaces and use the correlators but rather express the answer in terms of the matrix elements of this operator,
\begin{eqnarray}
\frac{\partial}{\partial t} S_M= 
\left(-M \sum_{a,\alpha;b,\beta} |H^{(AB)}_{a\alpha,b\beta}|^2 p^M_ap_\alpha \right.\\ \nonumber
\left.+M \sum_{a,\alpha;b,\beta}  |H^{(AB)}_{a\alpha,b\beta}|^2 p_bp_\beta p^{M-1}_a \right) 
\\ \nonumber
\int_{-\infty}^{t} dt' 2 {\rm Re} \left( e^{i(t-t')(E_i+E_\alpha -E_j-E_\beta)}\right)   
\end{eqnarray}
The integral over time $t'$ reduces to 
$$
2\pi \delta(E_a+E_\alpha-E_b-E_\beta),
$$
manifesting energy conservation between the initial state $|a\alpha>$ and final state$|b\beta>$.

This suggests that we can rewrite the whole expression in terms of Golden Rule 
rates $\Gamma_{a\alpha,b\beta}$ of the transitions between the states $|a\alpha>$
and $|b\beta>$,
\begin{equation}
\Gamma_{a\alpha,b\beta} = 2\pi |H^{(AB)}_{a\alpha,b\beta}|^2\delta(E_a+E_\alpha-E_b-E_\beta).
\end{equation}

With this, the flow reads
\begin{equation}
\left(S_M\right){\cal F}_M = 
M \sum_{a,\alpha;b,\beta} \Gamma_{a\alpha;b\beta}(p_b p_\beta-p_a p_\alpha)p^{M-1}_a 
\label{eq:2nd-general-form1}
\end{equation}
We see that the flow {\it vanishes} if the systems are in thermodynamic equilibrium at the same temperature. Indeed, in this case $p_b p_\beta / p_a p_\alpha = \exp((E_b+E_\beta-E_a-E_\alpha)/k_BT)=1$. 

Since the transition rates $\Gamma_{a\alpha,b\beta}$ in Golden rule approximation are symmetric with respect to a permutation $a\alpha \leftrightarrow b\beta$, we can regroup the terms  to arrive at
\begin{equation}
\left(S_M\right){\cal F}_M =  
M \sum_{a,b} \Gamma_{a\to b}p_a(p^{M-1}_b-p^{M-1}_a). 
\label{eq:2nd-general-form2}
\end{equation}
where 
$$
\Gamma_{a \to b} = \sum_{\alpha,\beta}  \Gamma_{a\alpha;b\beta}p_\alpha 
$$ 
gives the total transition rate from the state $|a>$ to the state $|b>$ averaged over all possible configurations of system $B$. Let us use Eq. (\ref{eq:2nd-general-form2}) to derive a simplified expression valid in zero-temperature limit. In this limit, the system $A$ is initially in the ground state $|0>$, so that $p_0=1$ and $p_a=0$ for $a \ne 0$, $S_M =1$.
We obtain 
\begin{equation}
{\cal F}_M = - M \Gamma_0;
\label{eq:2ndorder-zeroT}
\end{equation}
$\Gamma_0$ being the total transition rate from the ground state to any other state.
Remarkably, this involves no assumption concerning the system $B$: it can be very far from equilibrium.

Eq.  (\ref{eq:2nd-general-form2}) is also a convenient starting point to derive the expression for the flow of Shannon entropy $S$. Taking the limit $M \to 1$, we obtain 
\begin{equation}
-\frac{\partial S}{\partial t} =  \sum_{a,b} \ln\left(p_b/p_a\right)\Gamma_{a \to b} p_a. 
\label{eq:2nd-entropy}
\end{equation}
Let us assume thermal equilibrium of $A$. In this case, $\ln\left(p_b/p_a\right) = (E_a -E_b)/k_BT)$. Summing up the energy changes $E_b -E_a$ in the course of individual transitions from $a$  to $b$, we prove that the energy flow to the system $A$ equals
$$
\frac{d E}{d t} = \sum_{a,b} \Gamma_{a \to b} (E_b-E_a) p_a
$$  
Comparing this with Eq. (\ref{eq:2nd-entropy}), we recover the text-book relation between the heat and entropy flows
\begin{equation}
\label{eq:textbook}
\frac{d S}{dt} = \frac{1}{k_B T} \frac{d E}{dt},
\end{equation}
that appears to be universally valid within the second-order perturbation theory. Remarkably, this involves no assumption about the system $B$.

\section{Example: higher-order diagrams}
\label{higherorder}

Let us analyze the fourth-order diagrams for time derivative of $S_M$. As above, we assume that $H_{AB}$ does not contain diagonal elements. Since white contours are closed within each world, the four perturbations can either all come in the same world or in two pairs in two different worlds. If all four come in the same world, they describe a correction to one of the Golden Rule transition rates. This correction does not bring anything new and we disregard these diagrams in further consideration. 

\begin{figure}
\label{fig:fourth-order}
\includegraphics[width=0.35\textwidth]{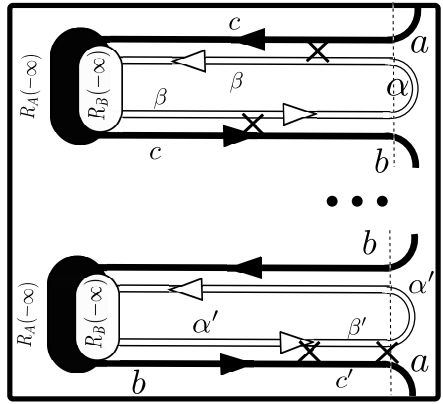}
\caption{ A fourth-order ''quantum'' diagram for R\'{e}nyi entropy flows.
The contributions come from perturbations $\hat{H}^{(AB)}$ in
two different worlds, only these two worlds are shown. The letters on the contours label the states involved.}
\end{figure}

A diagram involving two different worlds is given in Fig. \ref{fig:fourth-order}. We see that in general the black contour entering a world with perturbations exits it with a different state index. For a particular case when these indices are the same,   
$a =b$, the diagram diverges upon integration over time. This is not surprising since  we expand $S_M(t) \propto \exp ( {\cal F}_M t)$. The fourth-order expansion thus contains terms $\propto ({\cal F}^{(2)}_M)^{2} t/2$, ${\cal F}^{(2)}$ being the second-order contribution to the rate that we have already calculated. Indeed, the diagram with $a=b$ is proportional to $({\cal F}^{(2)})^2$ and therefore does not contribute to fourth-order correction to the flow. 
We thus concentrate on the case $a \ne b$. We call this diagram ''quantum'' since we will see that it does not permit an interpretation in terms of ''classical'' transition events.
All expressions for ${\cal F},dS/dt$ in this Section give fourth-order corrections to these quantities. 

There are 16 diagrams of this sort corresponding to the number of ways the pairs of $\hat{H}^{(AB)}$ in each world can be placed on bra and ket contours.
Summing up all of them, we can present the fourth-order correction in the following form:
\begin{eqnarray}
\label{eq:4th-general}
\frac{d}{dt} S^{(A)}_M = \pi \sum_{a,b} |A_{ab}|^2 \delta(E_a -E_b)\frac{p^{M-1}_a-p^{M-1}_b}{p_a-p_b};\\
A_{ab} = \sum_{c,\alpha,\beta} H^{(AB)}_{a\alpha,c\beta} H^{(AB)}_{c\beta,b\alpha} \nonumber\\
\left(\pi \left((p_a+p_b)p_\alpha - 2 p_cp_\beta\right) \delta\left(E_a+E_\alpha -E_c-E_\beta\right) - i\frac{p_a -p_b}{E_a+E_\alpha -E_c-E_\beta}\right).
\nonumber
\end{eqnarray}
The structure of the matrix elements in the ''amplitude'' $A_{ab}$ is the same as for an amplitude of the transition from the state $|a\alpha>$ to the state $|b\alpha>$, that is, without the change of the state of the subsystem $B$. Such transition would seem to involve a virtual state $|c,\beta>$. However, the rest of the expression for $A_{ab}$ does not support this interpretation: rather, probabilities enter in a form suggesting that the transition takes place between one of the states $|a\alpha>,|b\alpha>$ and the state $|c\beta>$. Therefore, the expression can be associated with no ''classical'' transition and corresponds to no actual transition rate.

Let us assume that the probabilities in the system $A$ depend only on energies of the corresponding states. Then it follows from $E_a=E_b$ that $p_a=p_b$. The term in $A_{ab}$ with the energy difference in the denominator vanishes and the flow reduces to 
\begin{eqnarray}
\label{eq:4th-f-of-E}
 S_M  {\cal F}_M= (M-1)\pi \sum_{a,b} |A_{ab}|^2 \delta(E_a -E_b) p^{M-2}_a;\\
A_{ab} = 2\pi \sum_{c,\alpha,\beta} H^{(AB)}_{a\alpha,c\beta} H^{(AB)}_{c\beta,b\alpha} 
\left(p_a p_\alpha - p_c p_\beta\right) \delta\left(E_a+E_\alpha -E_c-E_\beta\right). \nonumber 
\end{eqnarray}
We notice that if both systems are in thermal equilibrium, it follows from $E_a+E_\alpha =E_c+E_\beta$ that 
$p_a p_\alpha = p_c p_\beta$ and the ''amplitudes'' $A_{ab}$ vanish.

The ''quantum'' contribution derived manifests serious problems with term-by-term perturbation theory in the limit of vanishing temperature, indicating non-analytical dependence of the flows on coupling strength in the limit of weak couplings and vanishing temperatures. The contribution seems to have an evident zero-temperature limit, namely zero, at least if the ground state of the system $A$ is not degenerate. Indeed, delta-function in Eq. (\ref{eq:4th-general}) cannot be satisfied for any state $b\ne a$. However, analytical continuation to non-integer $M$ gives rise to problems.

To see this, we can attempt to derive the Shannon entropy flow taking the limit $M \to 1$ in Eq. (\ref{eq:4th-f-of-E}). We obtain 
\begin{equation}
\frac{d S}{dt} = \sum_{a,b} |A_{ab}|^2 \delta(E_a -E_b)\frac{1}{p_a};
\end{equation}
that is, the states with lesser probabilities $p_a$ contribute most to the entropy flow! Since the  probabilities of the excited states quickly decrease with decreasing temperature, we expect a divergence of the Shannon entropy flow at $T \to 0$ in contrast to vanishing Re-flows.

In \cite{Nazarov11}, the general expression has been elaborated for a typical quantum transport setup where the systems $A$ and $B$ are metallic leads kept at the same temperature but at different chemical potentials shifted by $eV$, and $\hat{H}_{AB}$ describes electron tunneling between the leads. The fourth-order Shannon entropy flow was found to diverge exponentially at $T \to 0$. This indicates an intriguing non-analyticity of the entropy flows in the coupling strength.

\section{Example: flows in quantum heat engine}
\label{QHE}
Let us give an example of computation of Re-flows in an interesting system.

A quantum heat engine (QHE) is a system of several discrete quantum states connected to the environments that are kept at different temperature. The motivation for research in QHE comes from studying models of photocells and photosynthesis. The thermodynamics of QHEs and their fluctuations in the quantum regime is not a continuation of classical results in discrete energies, instead features such as quantum coherence that have no classical analogue contribute to the heat exchange \cite{AN14}.

We consider a quantum system with discrete states $|n\rangle$  separated into two sets $\{u\},\{d\}$. All states within a set have approximately the same energy $E_{u}(E_{d})$,  the splitting $\epsilon_n$ within a set being much smaller than $E_u -E_d >0$. The system is subject to the external field with the frequency $\omega \approx E_u -E_d$ (we set $\hbar, k_B=1$ where appropriate) described by the Hamiltonian 
$H_{dr} = \sum_{m,n} \Omega_{mn} |m\rangle\langle n|  e^{-i\omega t} +H.c.$, and the relevant matrix elements are between the states of two sets. 
To distinguish the sets, let us introduce a matrix $\eta_{nm}$, $\eta_{nm} = 1$ if $n \in \{u\}$ and $m \in \{d\}$, $\eta_{nm} = -1$ if $n \in \{d\}$ and $m \in \{u\}$,
$\eta_{nm}=0$ otherwise.

The quantum system is coupled to a number of environments labeled by $a$ kept at different  temperatures $T_a$.  
We thus have a multi-partition: the whole space is separated onto the space of QHE states and the spaces of the environments.
The interaction with an environment is described by $H_{int}=  
\sum_{mn}|m\rangle\langle n|\hat{X}^{(a)}_{mn}$, with $\hat{X}^{(a)}_{mn}$
being the operators in the space of environment $a$. We assume linear response of each environment on the state of quantum system. In this case, each environment is completely characterized by the set of frequency-dependent generalized susceptibilities
$\chi^{(a)}_{mn,pq}(\nu)$ that are related to the correlators of {$\hat{X^{a}}$ defined as $S^{(a)}_{mn,pq}(t) \equiv \textup{Tr}_a \{\hat{X}^{a}_{mn}(0)\hat{X}^{a}_{pq}(t)\rho_a\}$}. The fluctuation-dissipation theorem yields the relations in frequency domain: {$S_{mn,pq}(\nu) =  n_B(\nu/T) \tilde{\chi}_{mn,pq}(\nu)$} where 
{$\tilde{\chi}_{mn,pq}(\nu) \equiv (\chi_{mn,pq}(\nu)-\chi_{pq,mn}(-\nu))/i$},
and  the Bose distribution ${n}_B(\nu/T) \equiv 1/ (\exp({\beta\nu})-1)$.

We concentrate on the Re-flows in one of the environments, which we call a probe environment. The rates induced by probe environment are assumed to be smaller than all other rates. In this case, we can concentrate on the second-order diagrams. We implement M-world Keldysh formalism where the contours of QHE and all environments except the probe one are closed within each world while the contour of the probe environment traverses all the worlds. There are two sorts of the second-order diagrams. The diagrams of the first sort, that we call incoherent, are within a single world and are similar to those considered in Section \ref{secondorder}. The presence of non-diagonal elements of density matrix in QHE gives rise to new type of diagrams, that we call coherent one. In this case, two perturbations are located in different worlds.

Collecting all diagrams (see Appendix B in \cite{AN14}), we obtain for ${\cal F}_M$ the following expression:
\begin{eqnarray}
&&{\cal F}_M = \frac{M n_B(M\omega/T)}{n_B((M-1)\omega/T) n_B(\omega/T) \omega} (Q_i - Q_c)
\end{eqnarray}

Thus the R-flow is naturally separated onto two parts, which come from  \emph{incoherent} and \emph{coherent} diagrams. The corresponding quantities $Q_{i,c}$ are expressed in terms of the density matrix of the engine $\rho$ and the dynamical susceptibilities of the probe environment,
\begin{eqnarray}
Q_i &=& \omega\   \sum_{mnp;\eta_{np}=1} 
\rho_{mn} \tilde{\chi}_{pm,np}(\omega)  (1+n_B(\omega/T))    -    \rho_{m{n}} \tilde{\chi}_{np,pm}(\omega) n_B(\omega/T)\    \\ 
Q_c &=&  \omega  \sum_{mnpq; \eta_{pq}=1} \rho_{nm} \rho_{qp} \tilde{\chi}_{mn,pq}(\omega)
\end{eqnarray}

The same-world diagrams contribute to the incoherent part that is proportional to $Q_i$. $Q_i$ is linear in $\rho$ so that is an observable. The different-world diagrams form the coherent part $\propto Q_c$ that is quadratic in $\rho$ and in principle would not be observable. The $M$ dependence is identical for both parts. 

Let us \emph{interpret} the parts and the quantities $Q_{i,c}$.  $Q_i$ is an observable: the total energy flow to the probe environment. The terms $\propto 1+n_B$ describe absorption of energy quanta $\hbar \omega$ by the environment, while those $\propto n_B$ correspond to the emission to the system. Upon taking limit $M \to 1$, the incoherent part reproduces the textbook equation for the entropy flow, ${\cal F}_S = Q_i/T_b$.

The interpretation of the coherent part is more involved and interesting. Let us replace  $|m\rangle\langle n|$ in  $H_{int}$ the operators with classical external forces {$f_{mn}$} with time-dependence {$f_{mn} \propto \exp(-i\omega\eta_{mn})$}. These classical forces would cause energy dissipation to the probe environment that is determined from the forces and the dissipative part of  susceptibility {$\tilde{\chi}$}. This energy dissipation is $Q_c$. 

Both parts of R-flows can be extracted from the measurement results, although in a different way. The entropy flow is not directly related to { energy flow}. Rather, 
\begin{equation}
{\cal F}_S = (Q_i - Q_c)/T_b
\end{equation}
 the difference is due to quantum coherent effects in our heat engine. Similar relation holds for the Renyi entropy flow in the low-temperature limit  
\begin{equation}
{\cal F}_M = M (Q_i - Q_c)/\omega
\end{equation}  
(this limit does not commute with $M\to 1$ since ${\cal F}_S$ diverges at low temperatures). In the absence of coherent effects, low-temperature 
R-flow is readily interpreted semiclassically \cite{Nazarov11} as number of events (in our case, $\hbar\omega$ {quantum} absorptions) per second in $M$ parallel worlds. With coherencies, such simple interpretation does not work since ${\cal F}_M$ can be negative \cite{AN15}.

\section{Example: exact correspondence}
\label{exact}
Another example of the Keldysh multi-world formalism is a relation which we derive for coherent and incoherent second-order diagrams in general time-dependent situation. This relation gives an exact correspondence between formally unphysical Re-flows and physical observables, namely, the full counting statistics of energy transfers considered in Section \ref{extended}.

As discussed in section \ref{why}, the Renyi entropies in quantum physics are considered unphysical,  or non-observable, due to their nonlinear dependence on  density matrix. Such quantities cannot be determined from immediate measurements;   instead their quantification seems to be equivalent to determining the  density matrix. This requires reinitialization of the density matrix between many successive measurements.  Therefore the flows of Renyi entropy between systems are the conserved measures of non-physical quantities. 

An interesting and non-trivial question is: Is there any relation between the flows of Renyi entropy and the physical flows?   An idea of such relation was first put forward by Levitov and Klich in \cite{LevitovKlysh}, where they proposed that the Shannon entropy flow can be quantified from  the measurement of full counting statistics (FCS) of charge transfers.   The validity of this relation is restricted to vanishing temperature and obviously to the systems where interaction occurs by means of charge transfer. In this section we present a relation which is similar in spirit, for details see \cite{AN15}.  

Let us consider two quantum systems $A$ and $B$. We assume that the system $A$ is infinitely large and is kept in thermal equilibrium at temperature $T$.   The system $B$ is arbitrary: it can encompass several degrees of freedom as well as infinitely many of those. It does not have to be in thermal equilibrium and in general is subject to time-dependent forces.  It is convenient to assume that these forces are periodic with period $\tau$. However this period does not enter explicitly in formulation of our result, which is also valid for aperiodic forces.  The only requirement is that there is a stationary limit of the flows of physical quantities to the system $A$. The stationary limit is defined by averaging the instant flow over the period $\tau$.  For aperiodic forces it is determined by averaging over sufficiently long time interval. 
  
The energy transfer is statistical. In section \ref{extended} we discussed the full counting statistics of energy transfers. The FCS of energy transfer in system $A$  during the time interval {$[0, \mathcal{T}]$} can be determined from Eq. (\ref{eq:HABEvolutionExtended}).  For quantification of the Renyi entropy flow we need to define an \emph{auxiliary} FCS of energy transfer.  The most {general } interaction Hamiltonian is $ \hat  H_{AB}=\sum_{n}\hat{A}_{n}\hat{B}_{n}$ with $\hat{A}_n$ being operators in the space of the system in thermal equilibrium, and  $\hat{B}_n$ being those in the space of the arbitrary system. Let us replace $\hat{B}_n$ with their {\it average} values $\hat{B}_n \to \langle \hat{B}_n \rangle$.  The resulting Hamiltonian is that of the equilibrium system subject to time dependent external forces. Those induce energy transfers to the system to be characterized by a FCS.  We discuss below possible physical realization of the scheme. So we have two FCSs. In the limit of long ${\cal T}$, their cumulant-generating functions(Keldysh actions) are proportional to the time interval, ${\cal S}_i(\xi) = - \bar{{f_i}}(\xi)$ (incoherent) and ${\cal S}_c(\xi) = - \bar{{f_c}}(\xi)$ (coherent), $\xi$ being the counting field of energy transfer to/from the system $A$.

Our  main result  is the following exact correspondence: 
\begin{equation}\label{eq. corres}
\bar{\mathcal{F}}_{M}^{(\beta)}/M  =   \bar{f}_i^{(M\beta)}(\xi^*)-\bar{f}_c^{(M\beta)}(\xi^*) , \ \ \ \ \ \xi^*=i \beta (M-1)
\end{equation}
which indicates that the Renyi entropy flow of the order $M$ to the system kept at temperature $T=1/k_B \beta$  is exactly equal to the difference of the FCS of incoherent and coherent energy transfers to the system kept at temperature $T/M$ at the fixed characteristic parameter $\xi^*$.  This relation  is valid in the limit of weak coupling, where the interaction between the systems can be treated perturbatively.

There is an obvious classical limit for the case where the quantum system B is considered to be classical. All operators $\hat{B}_n$ are numbers  corresponding to classical forces acting on the system in thermal equilibrium. In this case the dynamics of the system is governed by the Hamiltonian in degrees of freedom of the system and therefore will be unitary.  In this case there will be no entropy flow.  This can be separately understood only from looking into the FCS in the the correspondence (\ref{eq. corres}): in this case $\bar{f}_i=\bar{f}_c$.

The entropy/FCS correspondence (\ref{eq. corres}) allows us to quantify the time flow of Renyi as well as Shannon entropy. These quantities are not accessible in direct measurement as they are non-linear functions of density matrix.  Direct measurements of density matrix for a probe environment requires characterization of reduced density matrix of an infinite system, which is a rather non-trivial procedure and needs the complete and precise reinitialization of the initial density matrix.  However, measuring the entropy flow from the correspondence requires that some generating functions are extracted  from determining statistical cumulants of transferred energy in experimental data. This can be done equally well for imaginary and real values of the characteristic parameter. The measurement procedures may be complex, yet doable and physical.

The correspondence can have many other advantages; for instance: a complete understanding of entropy flows may help to identify the sources of fidelity loss in quantum communications and methods to prevent or control them.

\section{Conclusions}
\label{conclusions}
We have formulated and illustrated here a fascinating extension of Keldysh formalism on multiple parallel worlds.
Keldysh contours in this scheme are different for different sub-parts of a quantum system, this provides dependencies between the worlds. We explain that the formalism naturally arises in the context of characterization of the flows of conserved measures: R\'{e}nyi entropies, and illustrate its similarities with single-world extensions of Keldysh formalism.

It is a big honour for us to present these results in a special issue celebrating numerous scientific merits of Leonid Veniaminovich Keldysh. We gladly appreciate his pioneering research that provided a powerful and indispensable tool for many generations of quantum physicists, us including, and wish him many happy returns of the day.

\begin{acknowledgments}
The research leading to these results has received funding from the European Union Seventh Framework Programme (FP7/2007-2013) under grant agreement n° 308850 (INFERNOS).
\end{acknowledgments}

\end{document}